# Enhancing Clinical Classifier Robustness with Deep Representation Learning Based OOD Detection on DNA Methylation Data

Md Zahangir Alom, Quynh T. Tran, Breuer Alexandar, and Brent A. Orr*

***Abstract*—**DNA methylation (DNAM) array-based tumor classification has established itself as a powerful clinical diagnostic tool, with machine learning models achieving high performance accuracy to classify a wide range of tumor types. However, these models are trained on fixed reference cohorts and can produce faulty predictions when applied to tumor types which are not represented in the training distribution, a critical vulnerability for clinical deployment. To address this, we propose the first end-to-end deep representation learning-based out-of-distribution (OOD) detection framework expressly designed for DNAm array data. Deep neural network (DNN) classifiers were trained on publicly available sarcoma (SARC) (GSE140686) and brain tumor (BT) (GSE90496) cohorts, and feature vectors were extracted from the bottleneck layers. Five OOD methods including 1dimension (1D)-Subspace Analysis (1D-SSA), K-Nearest Neighbors (KNN), Ridge Regression, Mahalanobis Distance (MD), and Sum of Squared Distances (SSD) were explored with 200 and 500-dimensional embeddings. Spectral discrepancy was used to discriminate in-distribution (IND) from OOD samples. Performance was measured via cross-validation and independent testing using holdout samples and an independent clinical dataset. We found 1D-SSA was the most effective approach applied to this task. In the sarcoma-IND/brain tumor-OOD setting, the proposed framework achieved an AUROC of 0.975 and True Negative Rate (TNR)@95TPR of 0.930, and an AUROC of 0.9569 and TNR@95TPR of 0.8660 for the brain tumor-IND/sarcoma-OOD setting. On the independent clinical test set, the method yielded a mean AUROC of 0.976, a DE of 0.032, an FPR@95TPR of 0.065, and a TNR@95TPR of 0.935, confirming strong generalizability to real-world clinical samples. These results show deep representation-based out-of-distribution detection as a practical, principled quality-control layer for clinical-grade methylation-based tumor classification systems.

## I. INTRODUCTION

Due to the remarkable success of deep learning (DL) models, they have become essential problem-solving tools in Bioinformatics and computational pathology. Deep learning models demonstrate state-of-the-art performance in image classification [1], gene expression profiling [2, 3], and clinical diagnostics [4]. These models also can extract meaningful representations from complex biological data [5]. DNA methylation (DNAM) arrays have emerged as a powerful platform for clinical tumor classification. Models built on differentially methylated probes from reference tumor cohorts can be applied to diagnostic samples achieving high performance accuracy, across diverse biological contexts [6-9]. The standard classifiers report a calibrated score (e.g., random forest (RF) vote fraction or deep neural network (DNN) SoftMax probability as an indirect confidence indicator. However, this score does not reliably indicate whether a sample is truly in the distribution of the reference data. In rare instances, a sample representing a tumor type outside the training data may still receive a high-confidence prediction due to overconfident extrapolation of the model outside their training domain, a particularly dangerous failure mode in clinical diagnostics.

Classical machine learning approaches assume that the training and test data are independent and identically distributed (i.i.d.), a condition known as in-distribution (ID) learning [10]. One of the main challenges in deploying robust AI models for clinical applications is ensuring that models do not blindly predict based on unfamiliar or non-representative inputs. Shifts in the distribution of test examples relative to the training distribution, a phenomenon known as out-of-distribution (OOD) testing, can yield inappropriate classification results [10]. In the clinical environment, the consequences of misclassification may be significant as it may result in inappropriate subsequent therapy or alter clinical surveillance. Recent studies have emphasized that the reliable detection of out-of-distribution (OOD) inputs is increasingly recognized as a precondition for deploying machine learning systems for critical applications [11].

OOD sample detection offers substantial benefit when used in biomedical applications, where model reliability can have profound clinical and scientific implications. First, it can serve as a safeguard against false biomarker discovery and its

Aluthors of this paper are affiliated with St. Jude Children's Research Hospital, Memphis,TN 38105 USA(e-mail: malom@stjude.org, qtran@stjude.org, abreuer@stjude.org, and Brent.Orr@STJUDE.ORG). The corresponding author is Brent A. Orr, MD, PhD. He is now with the Department of Pathology, St. Jude Children's Research Hospital, Memphis, TN 38105 USA. (e-mail: author@lamar.colostate.edu).

Supplemental materials are added in the Appendix Section of the paper.

association with different diseases. Secondly, OOD detection helps identify batch effects or sample contamination. Thirdly, OOD detection ability helps ensure the generalization validity of biomedical machine learning models to external cohorts. Finally, an OOD-aware model adaptively allows for rejection of uncertain predictions, ultimately improving performance, transparency, and maintenance of regulatory compliance in clinical diagnostics.

Hence, developing robust OOD learning algorithms is essential for advancing computational medicine. We propose an OOD analysis method that leverages deep feature representations, applying quality control for test samples in addition to applying classifiers. The contribution of the proposed model can be summarized as follows:

- We develop and evaluate a novel OOD analysis method for DNA methylation data using the deep learned representations of brain tumor and sarcoma classification models.
- The number of probes (25K, 50K, 408K) and embedding dimensions (200, 500, 1K) were optimized for OOD detection.
- Five different OOD detection methods were evaluated for OOD analysis, including 1-D subspace analysis (1D-SSA) [17], Mahalanobis distance (MD) [12], K-Nearest Neighbors (KNN) [16], Ridge Regression (RR) self-reconstruction [25], and sum of squared distances (SSD) [23].
- Completely independent clinical samples, including those yielding false positive classification scores on clinically deployed DNA methylation-based models, were tested and showed promising performance in consistently identifying OOD samples.

The remainder of the paper is summarized as follows. Section II discusses the related works of this study. The methods are described in Section II. Section III demonstrates the results and discussion. Finally, section IV concludes the work with future directions.

## II. Related Works

OOD detection encompasses five related sub-topics [11]: anomaly detection (AD), novelty detection (ND), open set recognition (OSR), and outlier detection (OD). All define an in-distribution representation, with the common goal of identifying deviating test samples. Yang *et al.* [11] provide a comprehensive survey of these methods, noting that the majority have been proposed and evaluated within imaging domains, leaving non-imaging modalities, including molecular data, largely unexplored. While generalized OOD detection has emerged as a well-established research field, however, most existing studies have focused primarily on visual data.

Several OOD methods leverage deep representation learning to characterize the in-distribution examples. The Mahalanobis distance (MD) [12] measures how far each test sample lies from the in-distribution feature mean using a covariance structure and is an effective multivariate anomaly metric with demonstrated utility in imbalanced classification and one-class classification settings [REF?]. Hendrycks and Gimpel [13] established a baseline using prediction probability (PP) distributions from the neural network SoftMax layer, observing that in-distribution samples consistently yield higher maximum SoftMax probabilities than OOD samples, a simple yet effective signal for OOD detection. Building on probabilistic foundations, Haroush *et al.* [14,15] framed OOD detection as a statistical hypothesis testing problem, deriving per-sample p-values and providing formal control over the Type I Error rate (T1E)—the rate of incorrectly flagging in-distribution samples as OOD. Sun *et al.* [16] subsequently introduced a non-parametric alternative: computing the K-nearest neighbor (KNN) distance between the deep embedding of a test input and those of the training set, applying a threshold to determine OOD status without any distributional assumptions. Together, these methods illustrate a progression from shallow statistical metrics to principled, deep, representation-based OOD detection, each exploiting a distinct property of the learned feature space.

More recent work has advanced deep representation learning for OOD detection through structured geometric and contrastive approaches. The union of 1-dimensional (1D) subspaces method [17] exploits a compact representation: OOD samples are less likely to overlap with known training classes, and the first singular vector provides a robust, hyperparameter-free representation without requiring OOD samples during training for image classification problems. Energy-based models (EBMs) [18] define an energy function over inputs, flagging high-energy inputs as OOD. Self-supervised outlier detection (SSD) further leverages the geometry of the deep feature space by measuring deviations from the training distribution without requiring class labels. Contrastive learning [19] has also been explored to learn tighter class-conditional representations, demonstrating that representations optimized for intra-class compactness and inter-class separability yield improved OOD detection discovery. These methods spanning subspace geometry, energy landscapes, and contrastive objectives represent the current state of the art in OOD detection, yet all have been developed and validated exclusively in computer vision or natural language processing settings.

Despite this progress, the application of OOD detection to genomic data remains critically underexplored. The comprehensive survey by Yang *et al.* [11] and the imaging-focused review by Hong *et al.* [20] confirm that virtually all evaluated methods target computer vision tasks. In the biomedical domain, OOD detection has been investigated for radiology and other medical imaging problems [20], but no prior work has specifically addressed OOD detection for molecular datasets derived from DNA methylation arrays. High-performing DNA methylation classifiers for brain tumors [6], sarcomas [7], and other tumor entities [8,9] have demonstrated high levels of diagnostic accuracy and have been deployed in the clinical environment. Despite clinical deployment, methods to incorporated principled OOD detection mechanisms have not been established, leaving no reliable means to flag samples that fall outside their training distribution.

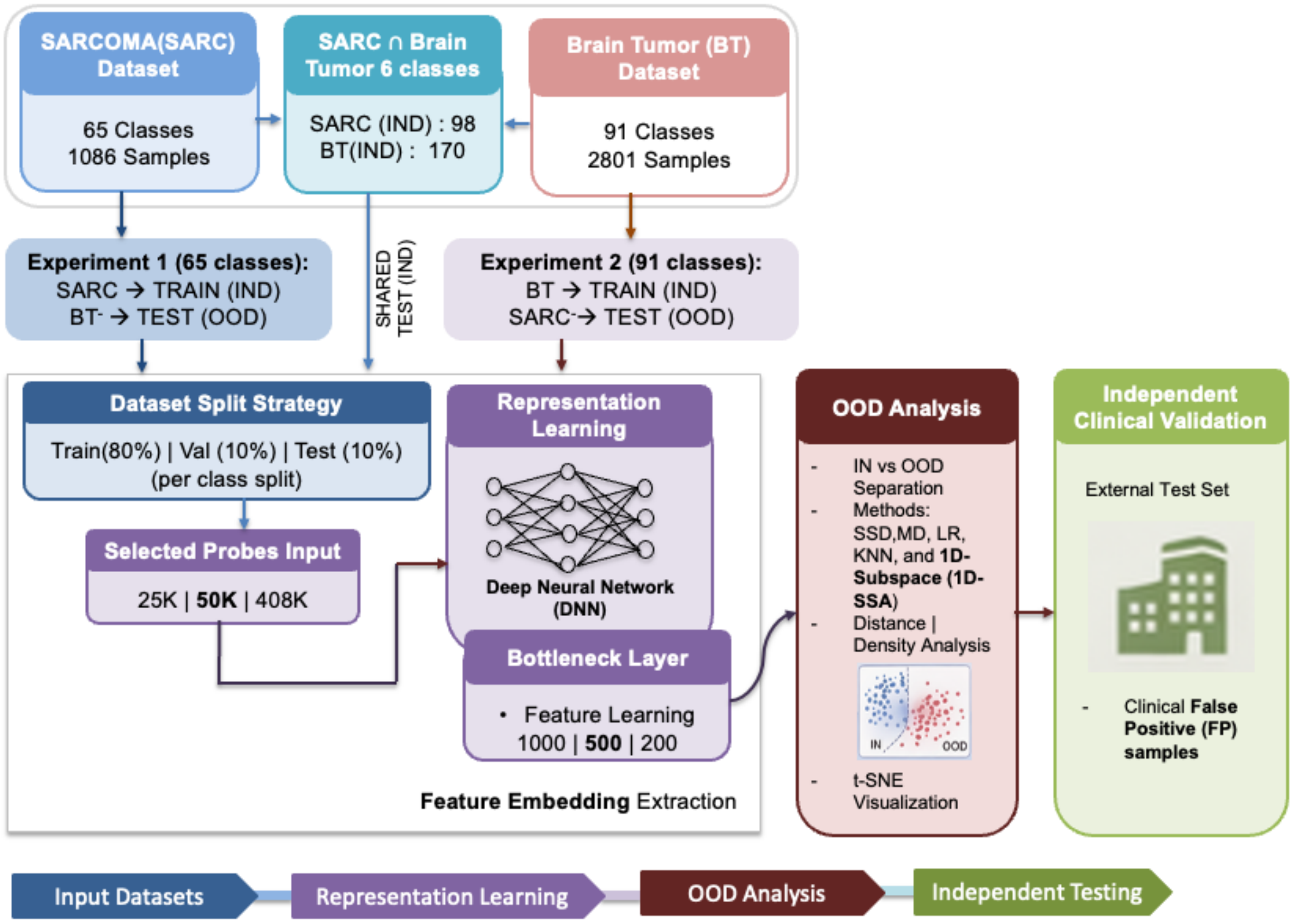

Fig. 1. The schematic diagram of the out-of-distribution (OOD) pipeline analysis systems consists of data preparation, learning and representation, OOD analysis, and visualization phases. We used a similar pipeline in both the SARC (65 classes) and the BT GSE-91 (91 classes) datasets. We conducted two sets of experiments. First, we considered SARCOMA as the training (IN-distribution) and GSE-91 (OUT-distribution) for testing in OOD analysis. We considered GSE-91 (IN-distribution) for training and SARC (OUT-distribution) for testing the models in the second setup. The total samples were 1086 and 2801 for the SARC and BT datasets, respectively. For both datasets, 80 percent of examples were used for training, 10 percent for validation, and 10 percent for testing, with equal numbers per class. We trained the DNN model using selected probes 25K, 50K, and 408K (after discarding sex-related probes) for the SARCOMA-based OOD detection system. The deep feature representation from the DNN models' bottleneck layers was used for out of distribution analysis (ODA). We evaluated performance across different feature embedding spaces (500 and 200). Finally, we evaluated the proposed models with independent clinical datasets.

This proposed work directly addresses that critical gap by adapting and evaluating deep representation learning-based OOD detection methods for clinical DNA methylation data, a domain where reliable uncertainty quantification has direct patient safety implications

## III. Methods

### *A. Datasets*

In this study, two publicly available reference datasets (one for brain tumor classification and one for sarcoma classification) were used to develop and validate OOD detection models.

**Sarcoma Dataset (SARC / GSE140686):** 1,086 soft tissue and bone tumor samples representing 65 classes, profiled on Illumina EPIC arrays [7]. Training: 832 samples; validation: 116; holdout test: 129. For testing on BT datasets, common cross-class/samples (6/98) between SARC and BT were treated as IND, and the remaining BT samples were treated as OOD examples.

**Brain Tumor Dataset (BT / GSE90496):** 2,801 CNS tumor samples representing 91 methylation classes, profiled on Illumina 450K arrays [6]. Training: 2,204 samples; validation: 320; holdout test: 277. For testing on SARC datasets, the common class/samples (6/170) were treated as IND, and the remaining 907 samples as OOD.

Two cross-dataset OOD configurations were evaluated: (i) SARC as in-distribution (IND) with BT as OOD; and (ii) BT as IND with SARC as OOD. Prior to model development, six shared classes between the two datasets were identified and separated: 98 intersected examples in the SARC-IND/BT-OOD setting and 170 in the BT-IND/SARC-OOD setting. These shared-class (CONTR-REACT/CTRL-REA, EFT-CIC/SBRCT-CIC, SCHW/SHW, SFTHMPC/SFT, CHORDM/CHORD, EWS) samples were treated as IND within the OOD test set and represent important internal IND controls.

**Independent Clinical Samples:** for investigating the clinical reliability of the proposed model, we collected 33 (from EPIC: 27 and EPICv2: 6) independent false positive clinical samples, representing tumor examples which were known to be outside the training data and where the deployed clinical model [21] failed to yield accurate class calls.

The reported performance was computed based on the test (IND) and test (OOD) examples for both experimental setups, respectively. For independent validation, test (IND) and independent test (OOD) samples were used, and the results were reported accordingly.

### *B. Deep feature representation*

DNN classification models were trained on each dataset using selected differentially methylated probes of 25K, 50K, or 408K (after dropping sex-related probes and probes with known mishybridization) from the DNA-methylation array. The DNN architecture consists of multiple fully connected layers with ReLU activations, batch normalization, and dropout regularization. The bottleneck layer—a layer of reduced dimensionality—serves as the source of deep feature representations for OOD analysis. Two embedding dimensions were evaluated: 200 and 500. All training used Adam optimization and early stopping on validation loss. Feature representations were extracted from the bottleneck layer for all training, validation, holdout test, and independent test samples. The end-to-end OOD analysis pipeline is illustrated in Fig. 1, and the DNN architectures are shown in Supplementary Fig. 1.

### *C. OOD detection*

Five different OOD detection methods were explored using deep feature representation from the bottleneck layers of DNN models.

**Union of 1-dimensional (1D) subspaces (UDS/1D-SSA)**: Following Zaeem Zadeh *et al.* [17], the first singular vector (FSV) of each training class was computed via singular value decomposition (SVD) of bottleneck features. The proposed **1D-SSA** provides robust representation for several reasons. First, the FSV is robust to perturbations and noise. Secondly, no OOD samples are needed in advance for model training, and it is applicable to any modality. Hence, we introduced an OOD detection method for the first time, using the first singular vector of the 1D (1D-subspace) derived from the deep-learned features from DNAM inputs. At test time, a sample was detected as OOD if its feature vector did not align with any learned singular vector, and the spectral discrepancy was high. This method required no hyperparameter tuning and no OOD samples during training, making it ideally suited for clinical deployment.

**The Mahalanobis distance (MD) OOD detection** measures how far each test sample is from the in-distribution feature mean, using the covariance structure of the training data [22]. The MD measures how far each test sample is from the center of the training distribution using the robust covariance. The class-conditional mean and covariance were estimated from training features; OOD scores for test examples were Mahalanobis distances. OOD samples scoring higher than range of Mahalanobis distance from reference samples were deemed OOD samples.

**Sum of Squared Distances (SSD) - OOD detection:** SDD-based OOD detection measures how far testing examples deviate from the training distribution in the deep-learned feature space [23, 25]. The mean vector of the training samples was computed, and the SSD scores were subsequently calculated as the sum of squared differences (squared Euclidean distance to the mean) for all test examples. Higher SSD scores indicate a greater likelihood that a sample is OOD [23]; whereas OOD samples lying outside the support of the training distribution will be geometrically isolated and yield substantially higher distances.

**Ridge Regression (Self-Reconstruction) (RRSR)**: We trained the model to learn and reconstruct the input deep feature representation. Then, compute the reconstruction error for in-distribution samples; they should reconstruct well, resulting in lower errors. The OOD samples reconstruct poorly with high error (OOD score). The reconstruction error serves directly as the OOD score [25]. Since all features are equally important in the lower dimension deep representation space, we applied a linear Ridge Regression (RR) model with (parameters: $\alpha = 1.0$) using scikit-learn [24]. The models were trained for 100 epochs to optimize. The mean squared difference between the input and predicted features was used to optimize the regression model.

**K-Nearest Neighbors (KNN)-OOD:** For each class in the training set, a KNN model is fitted over the bottleneck-layer feature embeddings using K=5 (the value was selected empirically using training samples), implemented via scikit-learn [24]. At inference time, the mean Euclidean distance from a test sample to its K nearest neighbors in the training feature space was computed as the OOD score [16,19]. The core intuition was that IND test samples will reside in dense, well-populated regions of feature space and thus exhibit low mean neighbor distances,

Statistical analysis was conducted and the p-values were calculated using the student's t-test implemented in SciPy 1.0 [26]. We characterized each distribution using the median and interquartile range (IQR) as robust descriptive measures and reported the full range (minimum to maximum) to capture overall spread and identify OOD examples. To further evaluate group separation, a one-sided Mann-Whitney U test was applied to compare DNN prediction probability distributions between IND and OOD samples, with violin and box plots used to visualize central tendency and variability across both groups.

### *C. Performance metrics*

Model performance was evaluated during cross-validation and independent testing using a combination of holdout test samples and independent testing samples. The ability to correctly identify samples as IND or OOD was assessed using the Area Under the Receiver Operating Characteristic (AUROC) curve, Detection Error (DE), FPT@TPRx, which measures FPT when the TPR is x (e.g., 0.95), and the TNR@TPRx score. The spectral discrepancies were also shown to demonstrate whether samples were in distribution (IND) or OOD. AUROC: Area Under the Receiver Operating Characteristic curve - overall discriminative ability. Detection Error (DE): Minimum misclassification probability = 0.5 × (FPR + FNR) at optimal threshold. Lower is better. FPR@95TPR: False Positive Rate at TPR = 0.95. Lower is better. TNR@95TPR: True Negative Rate at TPR = 0.95. A strong detector yields TNR@95TPR ≥ 0.90. Higher is better (↑). All metrics were computed using the scikit-learn package [24].

## IV. Results

We utilized the deep feature representations from the bottleneck layers of brain tumor and sarcoma classification models. To determine the optimal embedding space, we evaluated feature dimensions of 200, 500, and 1000. Empirically, a dimension of 500 yielded the best OOD analysis performance, and therefore, the remaining analysis was conducted based on a 500-dimensional embedding space. For optimal probe/embedding selection, we evaluated the performance of the UDS/1D-SSA, KNN, and RR methods.

### *A. SARC-IND and BT-OOD*

For the experiments, we discarded the common classes (98 cross-class intersected examples) from the brain tumor datasets. Our findings indicate that training, validation, and testing deep feature representations with 50K probes are more effective than with 25K and 408K probes. We also found that the 500-deep feature embedding space yields the best results in Table I. For the sarcoma IND and brain tumor OOD samples, the model achieves the best performance, with an AUROC of 0.975, DE of 0.057, FPR@95TPR of 0.069, and TNR@95TPR of 0.930, respectively. Using 408K (all) probes yielded substantially worse performance (AUROC 0.9469), while 25K (500-dim) was comparable to 50K.

TABLE I Performance of the UDS/1D-SSA OOD Models SARC Training (SARC-IND/BT-OOD): The symbols indicate the optimum direction of the score for each metric. Bold = best results

| Metric(Dir.) | 25K/ 200 | 25K/ 500 | 50K/ 200 | 50K/ 500* | 408K /200 | 408K /500 |
|---|---|---|---|---|---|---|
| AUROC(↑) | 0.969 | 0.975 | 0.973 | **0.975** | 0.947 | 0.946 |
| DE(↓) | 0.065 | 0.057 | 0.057 | **0.057** | 0.090 | 0.097 |
| FPR@95TPR(↓) | 0.085 | 0.093 | 0.069 | **0.069** | 0.130 | 0.151 |
| TNR@95TPR(↑) | 0.914 | 0.906 | 0.930 | **0.930** | 0.869 | 0.848 |

** Best configuration. Column headers: Probes / Embedding dimension.*

For comparison with other OOD methods under the same optimal configuration (50K, 500-dim): the Mahalanobis distance method yielded AUROC 0.6066, DE 0.2618, FPR@95TPR 0.4729, TNR@95TPR 0.5270; FPR@95TPR 0.3178, TNR@95TPR 0.6822; SSD yielded AUROC 0.9052, FPR@95TPR 0.3806; Ridge Regression yielded AUROC 0.9307, DE 0.1453, and KNN showed AUROC 0.9735, DE 0.0595. See Table III for a full cross-method comparison. Overall, the 50K/500 configuration achieved the best performance, while 50K/200 produced comparable results in Table II. In contrast, the 500-dimensional embedding showed the best performances when using 408K input probes.

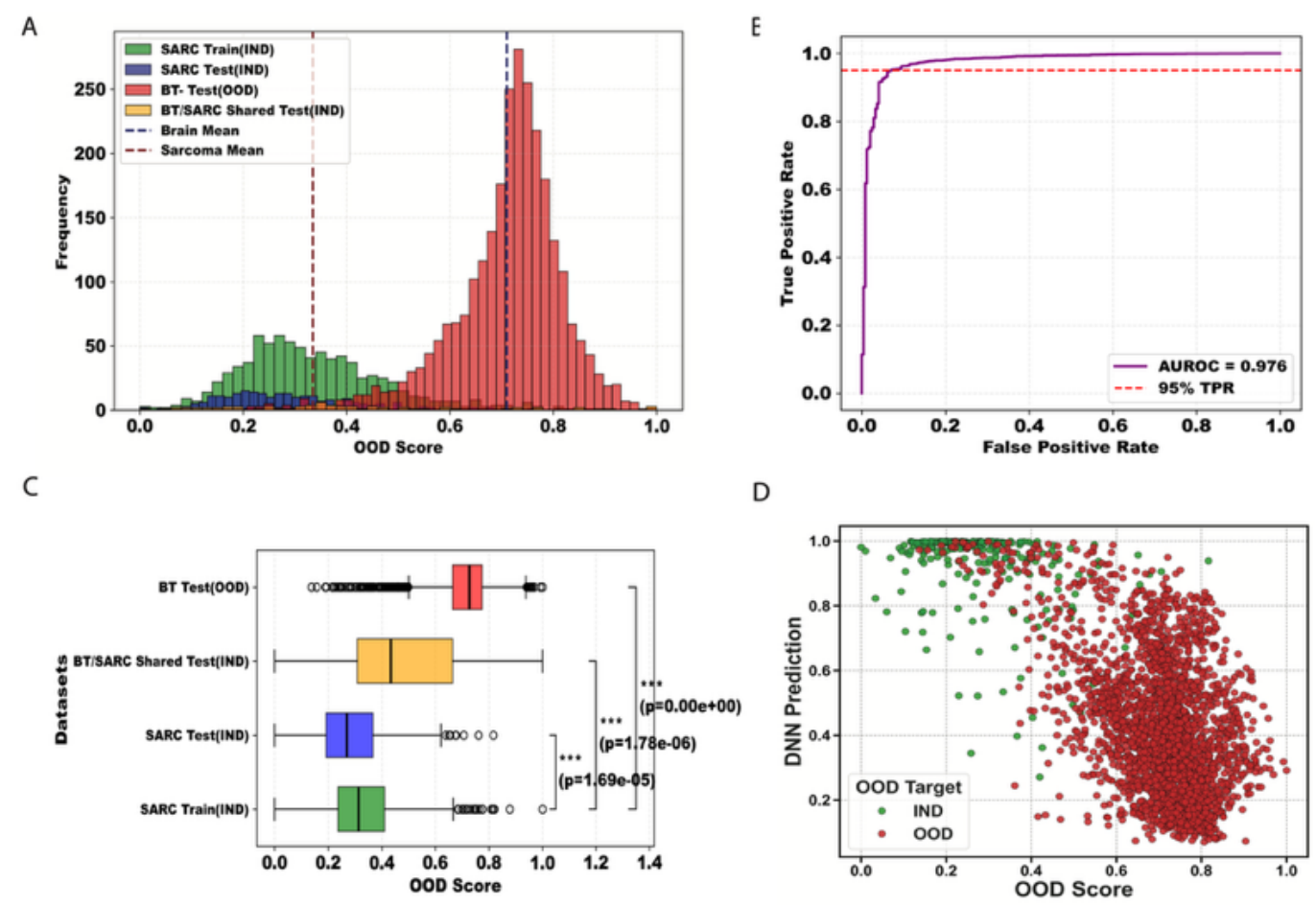

Fig. 2. The OOD analysis results were generated from the feature representations of the bottleneck layers of the DNN models, considering SARCOMA (IN) and Brain Tumor (BT) (OUT). For training the SARCOMA datasets, we used 832 cases for training, 116 for validation, and 129 for holdout test sets. The model was tested on 2703 cases from the BT(OOD) datasets. (A) shows the distribution of training and test examples. (B) shows the Area Under the Receiver Operating Characteristic curve (AUC-ROC) for this analysis, (C) The box plots with the p-value for test IND and OOD datasets with respect to the train sarcoma (IND) dataset, and, and (D) represents the DNN prediction with respect to the OOD predicted scores: 0 represents IND samples whereas 1 indicates OOD samples.

TABLE II Performance of the UDS/1D-SSA, KNN, and RRSR OOD Models BT Training (BT-IND/SARC-OOD): The symbols indicate the optimum direction of the score for each metric. Bold = best results

| Metric (Dir.) | UDS/1D-SSA | | KNN-OOD | | RRSR | |
|---|---|---|---|---|---|---|
| | 50K / 200 (Emb)* | 408K / 500 (Emb) | 50K / 200 (Emb) | 408K / 500 (Emb) | 50K / 200 (Emb) | 408K / 500 (Emb) |
| AUROC (↑) | **0.9660** | 0.9200 | 0.9551 | 0.9506 | 0.7887 | 0.8686 |
| Detection Error (DE)(↓) | **0..0787** | 0.1453 | 0.1105 | 0.1164 | 0.2706 | 0.1930 |
| FPR@95TPR (↓) | **0.1218** | 0.2968 | 0.2031 | 0.2094 | 0.5781 | 0.4062 |
| TNR@95TPR (↑) | **0.8781** | 0.7031 | 0.7969 | 0.7906 | 0.4219 | 0.5938 |

** Best configuration. Column headers: Probes / Embedding dimension.*

Fig. 2(A) shows OOD score distribution plots for training, holdout, and independent test sets with configuration of 50K/500. Panel A demonstrates a clear separation between SARC training samples (low OOD scores) and BT test samples (high OOD scores). Fig. 2(A) also confirms that 72 shared-class samples are correctly assigned low OOD scores. Fig. 2(B) shows the AUC-ROC of proposed OOD methods for FPT@95TPR. The dashed red line denotes the 95% true positive rate (TPR) threshold, at which the false positive rate (FPR) is 0.069 and the true negative rate (TNR) is 0.930.

The horizontal box plots for all four groups (center line, median; box limits, interquartile range; whiskers, 1.5× IQR; points, individual outliers) are shown in Fig. 2(C), showing the low OOD scores (median = 0.3131, Q1-Q3 = [0.2359, 0.4097], IQR = 0.1738, and outliers: 24/832) for training SARC-IND samples, while BT test samples exhibit high OOD scores (median = 0.7271, Q1-Q3 = [0.6637, 0.7741], and IQR = 0.1104, and outliers: 144/2729). Pairwise Welch's t-tests against the SARCOMA training set confirm highly significant score elevation in the BT/SARC shared cohort ($p = 1.78 \times 10^{-6}$) and the brain tumor OOD cohort ($p \approx 0$; ***), and a smaller but significant difference in the SARCOMA test set ($p = 1.69 \times 10^{-5}$). The t-SNE plot of DNN predictions versus OOD scores is shown in Fig. 2(D). For the sarcoma classification model, samples from the in-distribution set tend to receive the highest predictions from the DNN model, while OOD samples are assigned lower probabilities. The class-wise OOD score distributions (ordered by median) for SARC (IND) and BT (OOD) are shown in Supplementary Fig. 2. These results provide additional experimental evidence supporting the robustness of the proposed OOD analysis system.

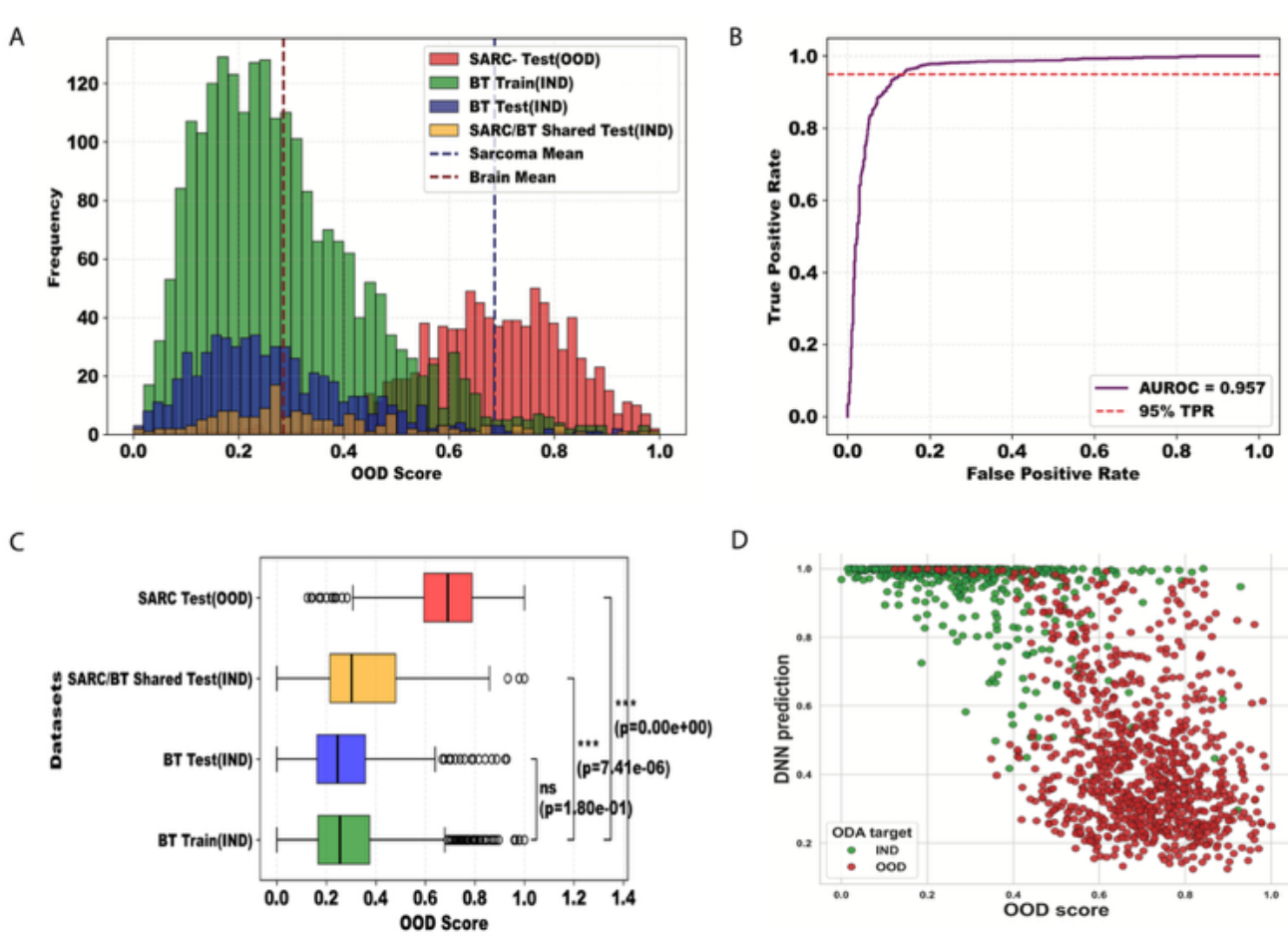

Fig. 3. The OOD analysis results for the DNN models considering Brain Tumor (IND) and SARCOMA (OOD). For training Brain Tumor datasets, we used 2204 cases for training, 320 for validation, and 277 for holdout test sets. The model was tested with 907 cases from the SARC(OOD) datasets: (A) represents the spectral distribution plot for training, holdout, and independent test sets. (B) ROC curves with AUC testing performance for Brain Tumor test examples (IN) in the SARCOMA test-out datasets. (C) The box plots show the distributions of the training and test sets for IND and OOD examples. Also shows the p-value for test IND and OOD datasets with respect to brain tumor train (IND) samples, and (D) represents the relationship between DNN prediction to OOD predicted score, ODA targets IND and OUT distribution examples.

### *B. BT-IND and SARC-OOD*

The BT-IND/SARC-OOD configuration used 2,204 BT samples for training (320 validation, 277 holdout test) and 907 SARC samples (excluding 170 shared-class examples) as OOD. Table II reports results for the 50K/200, and 408K/500 probe configurations. The model (50K, 200-dim) achieved an AUROC 0.957, detection error of 0.0872, FPR@95TPR of 0.1340, and TNR@95TPR of 0.8660, outperforming both the KNN (AUROC 0.9506, DE 0.1164, FPR@95TPR 0.2094, and TNR@95TPR 0.7906) and RRSR methods. The best model (50K, 500-dim) achieved AUROC 0.9569, DE 0.0872, which are shown in Table IV. The 408K configuration was substantially lower (AUROC 0.9200) for 408K/500 configurations shown in Table II.

TABLE III COMPARATIVE OOD DETECTION PERFORMANCE: WE COMPARE THE RESULTS FROM MD, SSD, RIDGE REGRESSION, KNN, AND 1D-SUBSPACE (UDS) METHODS (SARC-IND/BT-OOD, 50K PROBES, 500-DIM EMBEDDING). THE SYMBOLS INDICATE THE OPTIMUM DIRECTION OF THE SCORE FOR EACH METRIC. BOLD = BEST RESULTS.

| Method | AUROC ↑ | DE ↓ | FPR@95TPR ↓ | TNR@95TPR ↑ |
|---|---|---|---|---|
| Mahalanobis (MD) | 0.9099 | 0.2618 | 0.4729 | 0.5270 |
| Normalized MD | 0.9281 | — | 0.5530 | — |
| SSD | 0.9052 | — | 0.3806 | — |
| Normalized SSD | 0.8634 | — | 0.7069 | — |
| Ridge Regression | 0.9307 | 0.1453 | 0.3178 | 0.6822 |
| KNN | 0.9735 | 0.0595 | 0.0775 | 0.9225 |
| **1D-Subspace (UDS)*** | **0.9756** | **0.0576** | **0.0693** | **0.9306** |

For comparison, normalized Mahalanobis distance achieved AUROC 0.8431 and FPR@95TPR 0.6125 in this direction; Ridge Regression achieved AUROC 0.8538 and FPR@95TPR 0.5969; normalized SSD achieved AUROC 0.9008 and FPR@95TPR 0.3439

Fig. 3 illustrates the distribution of OOD scores across the training, holdout testing, and independent test datasets for configuration of 50K/500. BT(IND) training samples are clearly separated from the SARC(OOD) and BT(IND) testing examples overalled to the BT(IND) training examples in panel A. The dotted lines represent the median values for IND (red) and OOD (blue), respectively. Panel B presents the AUC-ROC curve for OOD analysis, and the model yielded an AUROC of 0.957. Panel C illustrates the box plots for IND and OOD samples, showing the low OOD scores (median = 0.2552, Q1-Q3 = [0.1663, 0.3743], IQR = 0.2080, and outliers: 54/2204), while SARC test samples exhibit high OOD scores (median = 0.6914, Q1-Q3 = [0.5948, 0.7884], IQR = 0.1936, and outliers: 13/917). Pairwise Welch's t-test for BT training set confirm highly significant score elevation in the SARC/BT shared cohort ($p = 7.41 \times 10^{-6}$) and the sarcoma tumor OOD cohort ($p \approx 0$; ***), and a smaller but significant difference in the BT test set ($p = 1.8 \times 10^{-1}$). Scatter plot of DNN maximum-class prediction confidence versus 1D-USP OOD score for IND (green circles) and OOD (red triangles) samples in D. IND samples cluster in the high-confidence, low-OOD-score quadrant, confirming concordance between classifier certainty and distributional membership whereas OOD brain tumor samples occupy high OOD scores across a wide range of DNN confidence values, demonstrating that DNN confidence alone is insufficient to flag out-of-distribution inputs. For BT (IND) and SARC (OOD), the class-wise OOD score distributions,

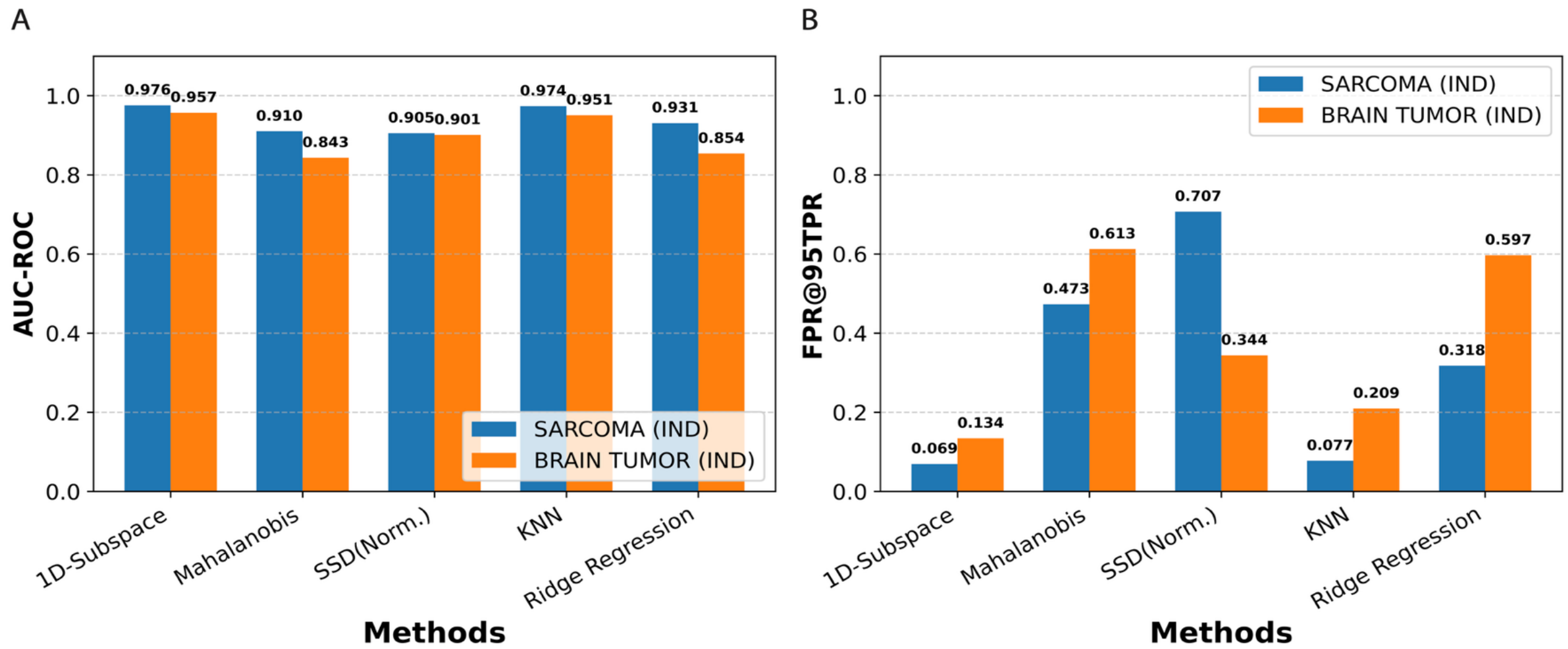

Fig. 4. Comparison of OOD detection performance across different methods: (A) AUC-ROC comparison- Bar plots showing the Area Under the Receiver Operating Characteristic curve (AUC-ROC) for five methods- 1D-Subspace, Mahalanobis, SSD(Norm.), KNN, and Ridge Regression on SARCOMA (IND) and Brain Tumor (OOD) datasets. The values are displayed at the top of each bar, with higher values indicating better discrimination. (B) FPR@95TPR comparison, Bar plots demonstrating the False Positive Rate at 95% True Positive Rate for the same methods and datasets, with values displayed on the top of the plots. Lower values indicate better OOD direction performance.

ordered by median per class, are shown in Supplementary Fig. 3.

### *C. Comparative Performance of OOD methods*

Table III provides a comprehensive comparison of all evaluated OOD detection methods for the SARC-IND/BT$^{-}$-OOD configuration at the optimal probe/embedding setting (50K, 500-dim). The 1D-subspace method (1D-SSA) achieved the best AUROC (0.9756) and specificity across all five methods. The Ridge Regression and K-nearest neighbors (KNN) methods showed moderate performance, while the SSD and Mahalanobis distance variants performed worse—consistent with known limitations of distance-based metrics in high-dimensional feature spaces.

Table IV presents a comprehensive comparison of all evaluated OOD detection methods under the BT-IND/SARC$^{-}$-OOD configuration at the optimal probe and embedding setting (50K probes, 500-dimensional embedding). Among the five methods, the 1D-subspace method (1D-SSA) achieved the highest AUROC (0.9569) and strongest specificity, demonstrating its effectiveness in discriminating OOD samples without requiring OOD examples during training. The statistical analysis shows the OOD score for BT-Test(IND) (median = 0.2461, Q1-Q3 = [0.1663, 0.3584], and IQR = 0.1958, outliers = 17/597) and (median = 0.6914, Q1-Q3 = [0.5944, 0.7885], and IQR = 0.1941, outliers = 13/907) for SARC-Test (OOD).

K-nearest neighbors (KNN) and MD (Norm.) delivered moderate performance, offering a reasonable balance between simplicity and discriminability. In contrast, SSD, Ridge Regression, and the Mahalanobis distance variants exhibited comparatively lower performance, a finding consistent with the well-documented degradation of distance-based metrics in high-dimensional feature spaces where covariance estimation becomes unreliable. In summary, the 1D-SSA demonstrated the highest testing performance for both datasets in terms of AUC-ROC and FPR@95TPR. The KNN and Ridge Regression methods showed moderate performance. The overall comparison across the methods is illustrated in Fig. 4.

TABLE IV COMPARATIVE OOD DETECTION PERFORMANCE: ALL METHODS MD, SSD, RIDGE REGRESSION, KNN, AND 1D-SUBSPACE (UDS) FOR BT-IND/SARC$^{-}$-OOD, 50K PROBES, 500-DIM EMBEDDING. THE SYMBOLS INDICATE THE OPTIMUM DIRECTION OF THE SCORE FOR EACH METRIC. BOLD = BEST RESULTS.

| Method | AUROC ↑ | DE ↓ | FPR@95TPR ↓ | TNR@95TPR ↑ |
|---|---|---|---|---|
| Mahalanobis (MD) | 0.8431 | 0.3305 | 0.6125 | 0.3870 |
| Normalized MD | 0.9558 | — | 0.1929 | — |
| SSD | 0.8538 | — | 0.5082 | — |
| Normalized SSD | 0.9008 | — | 0.3439 | — |
| Ridge Regression | 0.8538 | 0.2084 | 0.5969 | 0.4031 |
| KNN | 0.9506 | 0.1164 | 0.2094 | 0.7906 |
| **1D-Subspace (UDS)*** | **0.9569** | **0.0872** | **0.1340** | **0.8660** |

** Best practical method for clinical deployment. 1D-SSA requires no hyperparameter tuning and no OOD training samples.*

### *C. Independent Testing Performance*

The best SARC-trained 1D-SSA model (50K probes, 500-dim embedding) was applied to 33 independent clinical samples, representing samples receiving true positive and false positive clinical classification results. Table III summarizes performance. The model showed AUROC 0.9756, DE 0.0576,

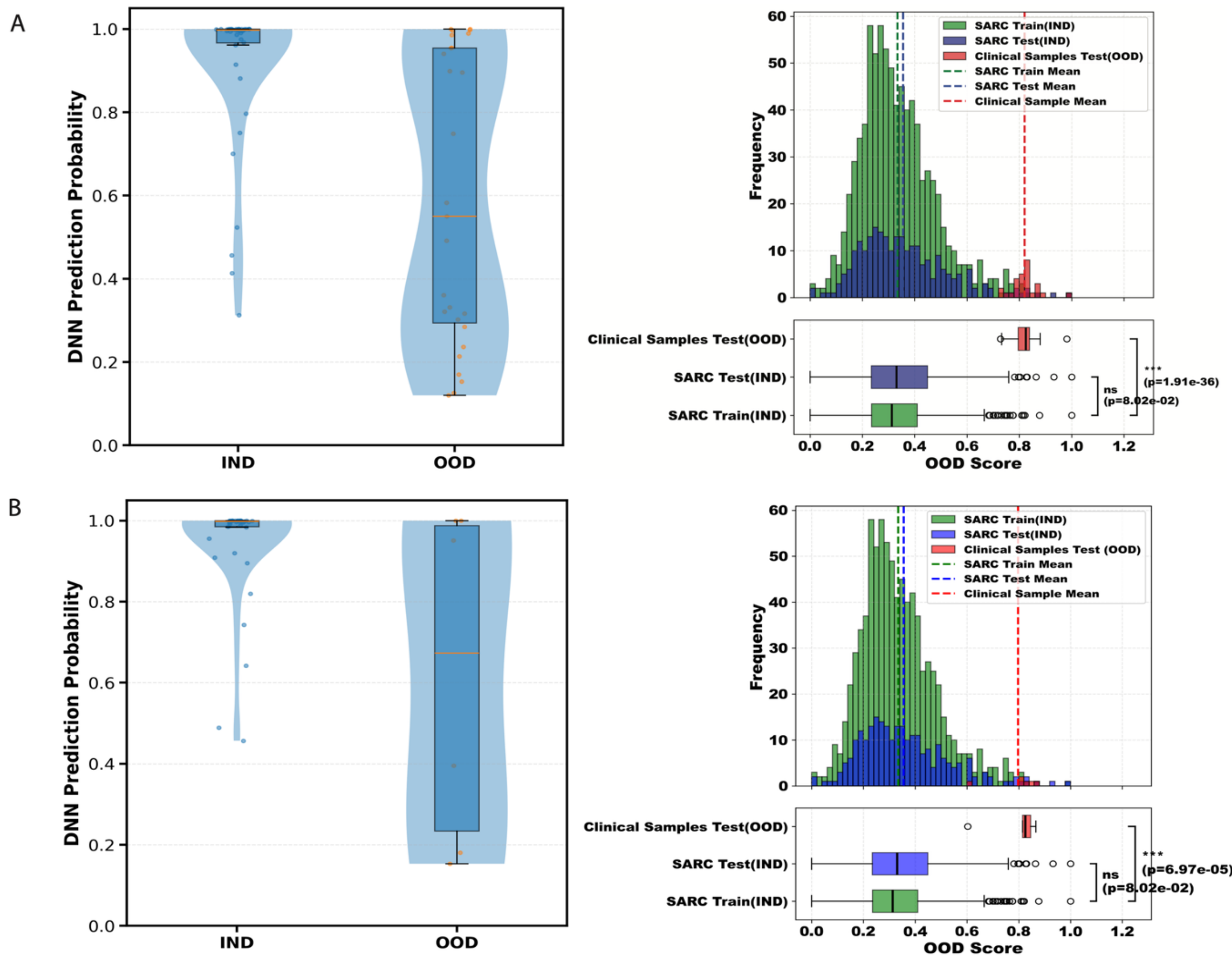

Fig. 5. Visualization of OOD detection for 33 independent clinical samples (EPIC (n=27) and EPICv2 (n=6) arrays): (A) illustrate the violine and box plots of DNN prediction IND and OOD samples (*p-value* < *0.05*). On the right side of panel A, Results for the EPIC array showing OOD detection performance. The distribution plot clearly shows that the test samples are well separated (*p-value* < *0.05*) and far from the Sarcoma IND training and clinical sample distributions. (B) Results for the EPICv2 array, demonstrating consistent OOD detection across six independent clinical cases.

FPR@95TPR 0.0693, TNR@95TPR 0.9306 with 50K probes; the 408K model was substantially worse (AUROC 0.9496).

We further validated the proposed best SARC-trained model (50K probes, 500-dim embedding) applied to 33 independent clinical samples. Fig. 5 illustrates the performance on 33 independent clinical samples derived from EPIC (n=27) and EPICv2 (n=6) arrays. The yielded AUROC of 0.9727, DE of 0.044, and FPR@95TPR of 0.0897 with 50K probes/500 feature embedding for the EPIC array.
The violin and box plots visualization on the left side of the pa nel in Fig. 5(A) confirm that the MLPNetv5[27] prediction pro babilities for OOD clinical samples (11/27) had high confidenc e (>0.9) compared to IND samples (randomly selected 40 sam ples from 'EWS', 'RMS, ALV', 'RMS, EMB', and 'OS, HG' clas ses). On the right side of the panel Fig. 5(A), results from the E PIC cohort demonstrate a clear separation between indiscretion Sarcoma training/testing samples and external clinical samples , as evidenced by the OOD score distributions. The statistical a nalysis indicates relatively low OOD score for SARC(IND) (m edian = 0.3131, Q1-Q3 = [0.2359, 0.4097], and IQR = 0.1738, outliers = 24/832), whereas the clinical testing samples exhibit substantially higher OOD scores (median = 0.8244, Q1-Q3 = [ 0.7955, 0.8393], IQR = 0.0437, and outliers = 2/27).
For the EPICv2 array, we achieved AUROC 0.9798, DE 0.020, and FPR@95TPR 0.0408 with 50K probes and 500-feature embedding. The violin and box plots demonstrate prediction scores between IND and OOD samples from MLPNet5 [27], where 3/6 samples were classified above the clinical threshold values > 0.9 On the right side of the panel Fig. 5(B) presents results from the EPICv2 array, showing that six independent cases exhibit a consistent pattern of elevated OOD scores, showing the low OOD scores (median = 0.3131, Q1-Q3 = [0.2359, 0.4097], and IQR = 0.1738, outliers = 24/832) for SARC, while clinical test samples exhibit high OOD scores (median = 0.8256, Q1-Q3 = [0.8149, 0.8459], IQR = 0.0310, outliers = 1/6). These results confirm the robustness and generalizability of the proposed model across the independent clinical datasets and array platforms.

### *D. Probe Selection and Embedding Dimensionality Ablation*

The 50K probe set consistently outperformed both 25K and 408K alternatives across all configurations, reflecting a signal-to-noise trade-off: 25K probes may miss informative differential methylation patterns, while 408K probes include noisy, uninformative loci that degrade representation quality. The 500-dimensional embedding consistently outperformed 200-and-1,000-dimensional embeddings for both tasks, providing an effective balance between representation expressivity and compactness for 1D subspace learning.

## V. Discussion

We demonstrate an end-to-end robust out-of-distribution (OOD) detection system utilizing deep representation learning from DNA methylation arrays. The experimental results show promising performance in OOD detection for different tumor classification datasets (demonstrated 0.975 and 0.959 ROC-AUC scores for holdout test sets from SARCOMA and Brain Tumor datasets using 1D-SSA). The framework generalizes to an independent clinical cohort (AUROC of 0.9798 for EPIC and 0.9727 for EPICv2) using 1D-SSA, validating its utility for real-world clinical quality control.

Among the evaluated methods, the 1D-subspace analysis (1D-SSA) demonstrated the best practical performance. Its key advantages for clinical use are: (1) no hyperparameter tuning required, (2) no OOD training samples needed for training, and (3) applicability to any data modality. The K-nearest neighbors (KNN) showed moderate performance (AUROC 0.973, Sarcoma-IND/BT$^-$-OOD), offering a computationally straightforward alternative when DNN feature alignment is not feasible. Distance-based methods (MD, SSD, and variants) showed lower performance (AUROC 0.86–0.93), consistent with known limitations of simple distance metrics in high-dimensional feature spaces. The normalized MD and SSD variants showed improvements, confirming that feature normalization is important in this domain.

The independent validation on known false positive clinical samples is a particular strength of this study. Performance on 33 prospectively collected clinical samples (mean AUROC 0.9762, DE 0.032, FPR@95TPR 0.0649, and TNP@95TPR 0.9346) demonstrates that the framework is robust to the variability encountered in clinical practice—including rare tumor types, unusual presentations, and samples outside the training distribution. Notably, the samples deemed in distribution on this set represented closely related tumor entities driven by similar driver abnormalities that are encountered both the brain and periphery. In addition, the testing results demonstrates the robustness across different methylation arrays including EPIC and EPICv2. Integration of OOD could represent an important quality control mechanism in the clinical environment. In distribution samples with classifying scores would provide additional confidence in reliability of classification performance. Samples flagged as OOD could protect against misreporting false positive classification in the clinical environment but also could prompt additional workup: subspecialty pathology review, orthogonal molecular testing (e.g., comprehensive genomic profiling), or molecular tumor board consultation.

Limitations include: (1) the requirement for an independently trained DNN classifier whose quality effects OOD detector performance; (2) primary focus on two large cohorts—performance on smaller specialized datasets remains to be confirmed; and (3) the optimal probe selection may vary based on reference tumor types and may need optimization as reference tumor types are updated. Future work will focus on prospective more comprehensive prospective clinical validation and clinical workflow integration.

## VI. Conclusion

We have developed and validated a deep representation learning-based OOD detection framework for quality control of DNA methylation array-based clinical tumor classification. Five OOD detection methods were systematically evaluated 1D-Subspace (1D-SSA), K-Nearest Neighbors (KNN), Ridge Regression, Mahalanobis Distance (MD), and Sum of Squared Distances (SSD); the 1D-subspace spectral discrepancy method achieved the best practical performance with AUROC scores of 0.976 (SARC-IND/BT$^-$-OOD), 0.957 (BT-IND/SARC$^-$-OOD), and 0.9762 on an independent clinical dataset of 33 St. Jude samples. The 50K probe / 500-dimensional embedding configuration consistently delivered optimal results. These findings establish OOD detection as a feasible and clinically valuable quality control component for methylation-based tumor classification workflows, supporting safer and more reliable AI-assisted diagnosis in oncologic pathology.

## Appendix

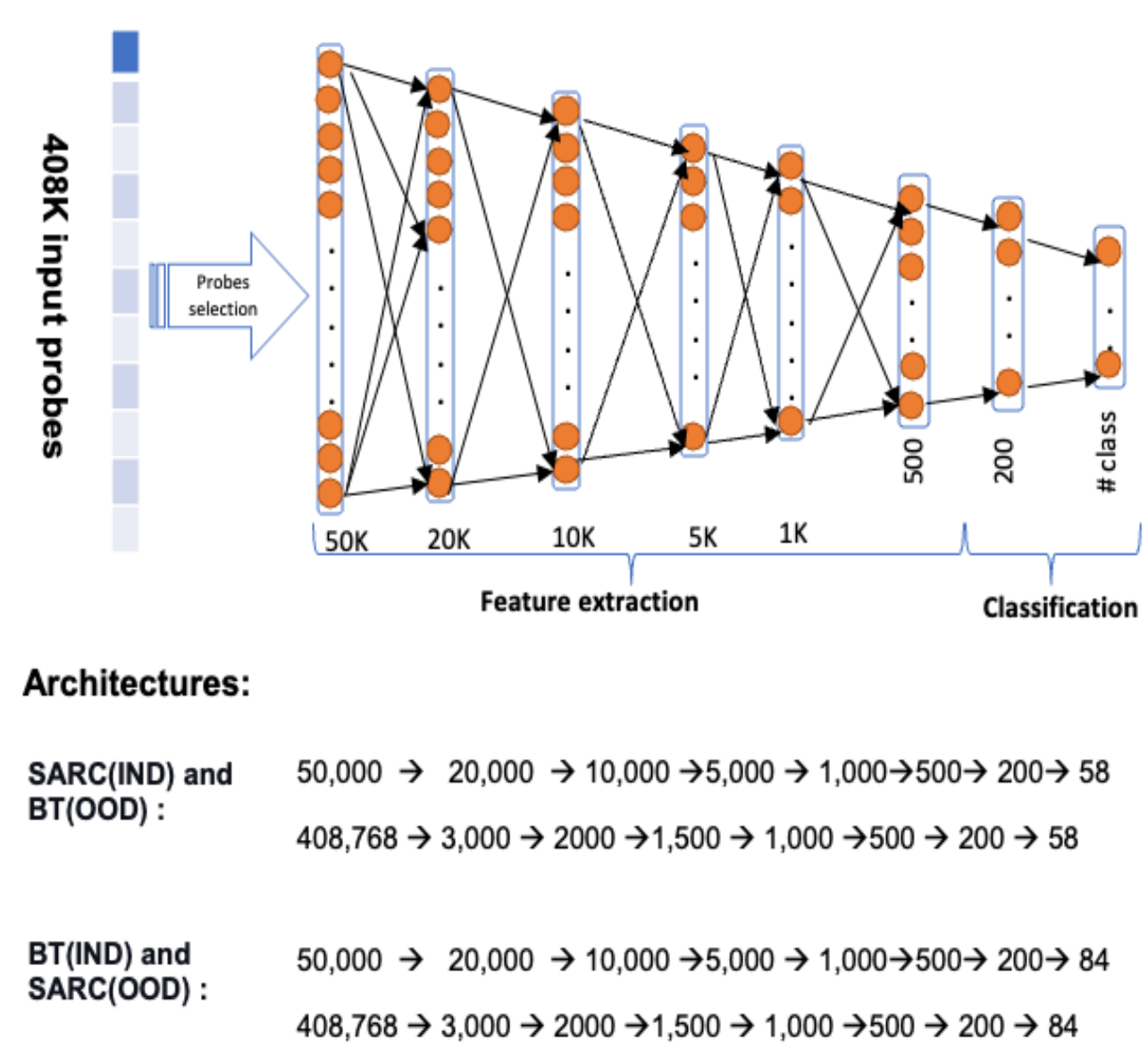

Supplementary Fig. 1. The Deep Neural Networks (DNN) architectures for both experimental setups: the DNN architecture for 50K input probes from 408K, and the architectures for 408K input probes.

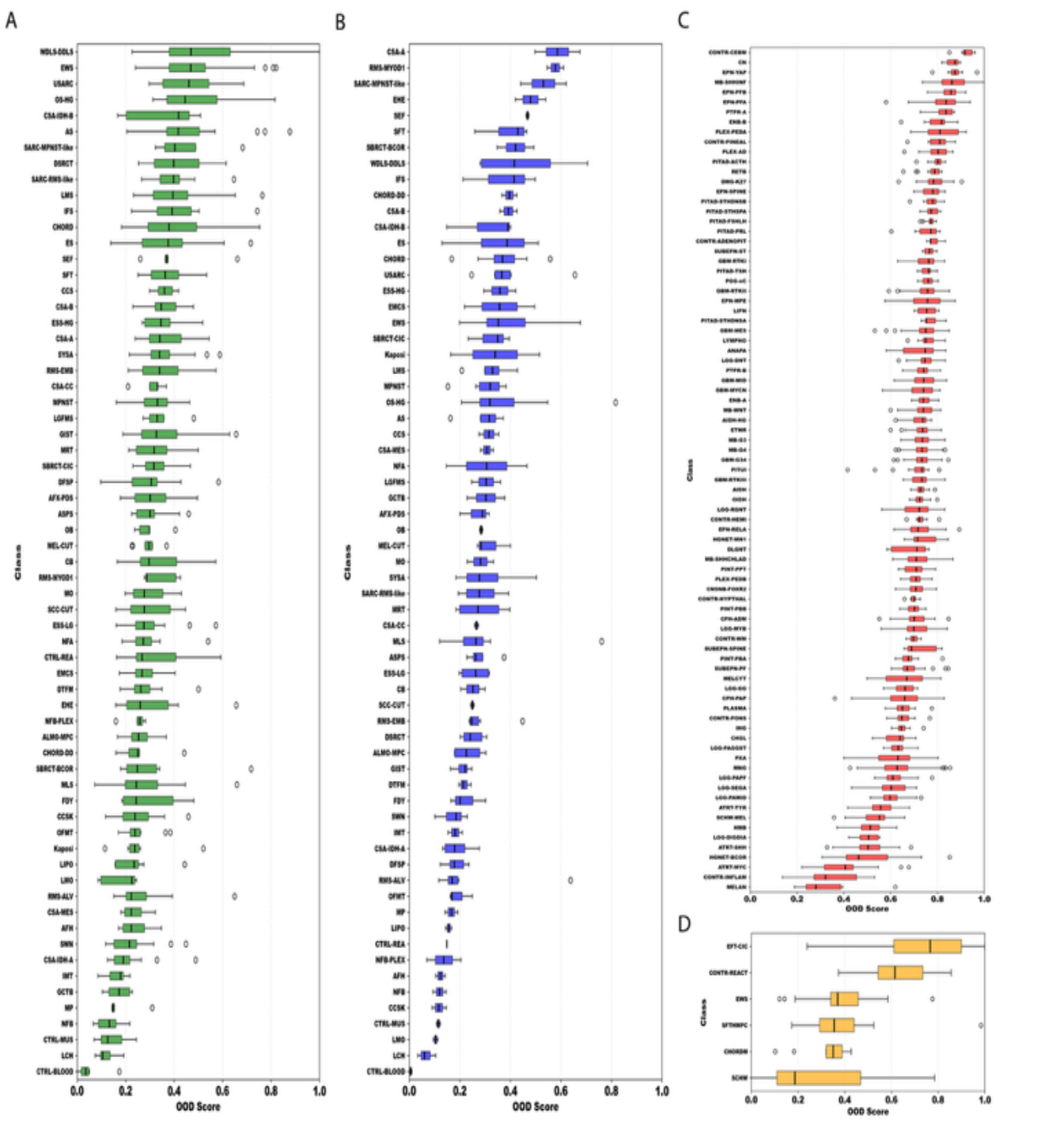

Supplementary Fig. 2. Class-wise OOD Score Distribution (Ordered by Median) of training and testing examples for SARC(IND) and BT(OOD): (A) and (B) show the class-specific box plots for SARC training (IND) and testing (IND) samples, respectively. (C) Class-wise box plots for BT (OOD) examples, which show a higher OOD score compared to IND samples, and (D) shows the box plots for IND samples from the Brain Tumor datasets.

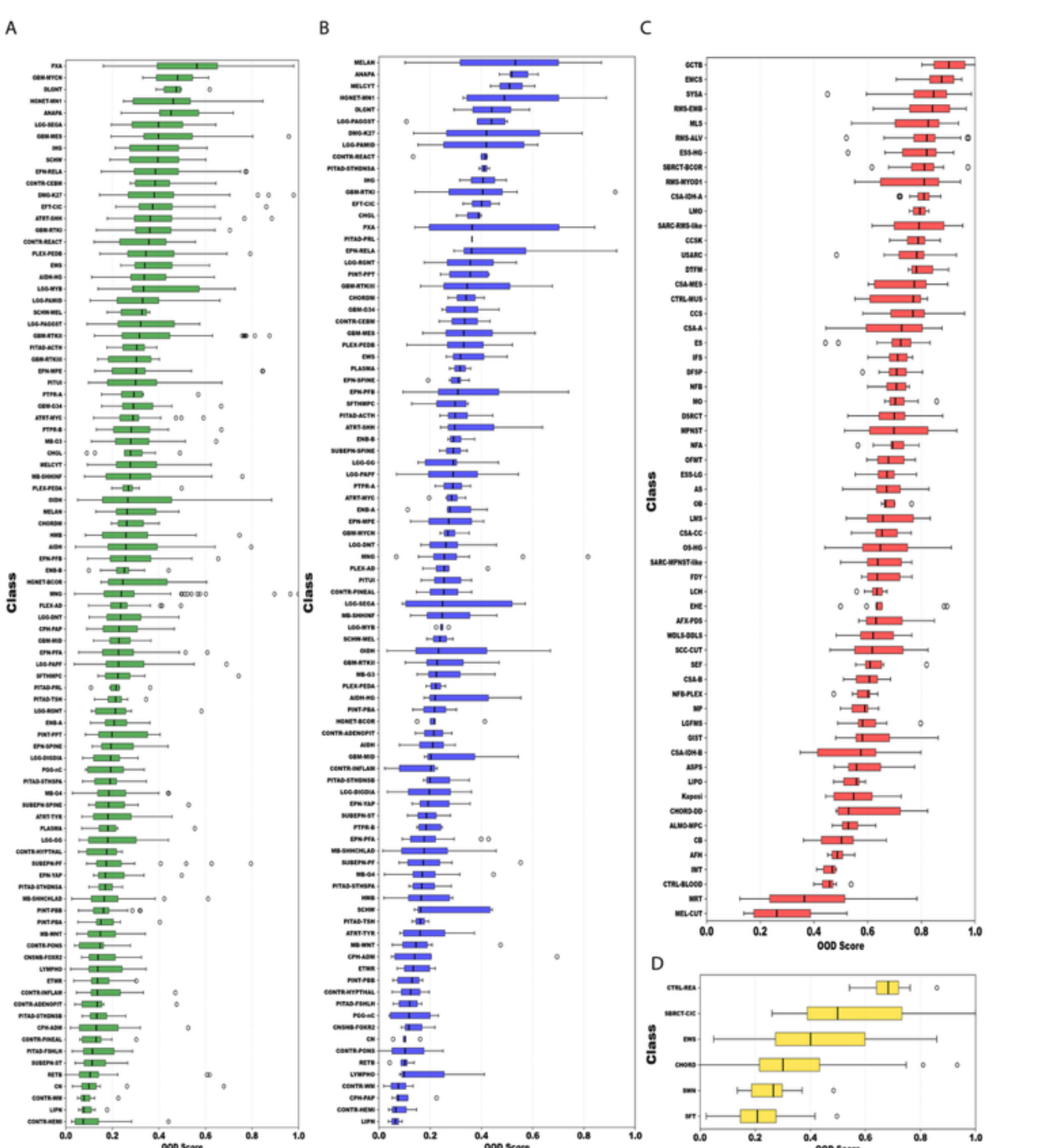

Supplementary Fig. 3. Class-wise OOD Score Distribution (Ordered by Median) of training and testing examples for Brain Tumor (BT)(IND) and SARCOMA(OOD): (A) and (B) show the class-specific box plots for BT training (IND) and testing (IND) samples, respectively. (C) Class-wise box plots for SARC (OOD) examples, which show a higher OOD score compared to IND samples, and (D) IND samples from the SARCOMA datasets.

## Acknowledgment

The authors thank members of the Department of Pathology at St. Jude Children's Research Hospital for clinical data contributions and helpful discussions. This work was partially supported by the National Cancer Institute support grant (P30 CA021765) and the American Lebanese Syrian Associated Charities (ALSAC). The authors acknowledge use of publicly available datasets from NCBI GEO (GSE90496 and GSE140686).

## References

[1] Krizhevsky A, Sutskever I, Hinton GE. ImageNet classification with deep convolutional neural networks. Commun ACM 2017;60:84–90. https://doi.org/10.1145/3065386

[2] Lee M. Deep learning techniques with genomic data in cancer prognosis: A comprehensive review of the 2021–2023 literature. Biology 2023;12:893. https://doi.org/10.3390/biology12070893

[3] Chafai N, Hayah I, Houaga I. *et al.* A review of machine learning models applied to genomic prediction in animal breeding. Front Genet 2023;14:1150596. https://doi.org/10.3389/ fgene.2023.1150596

[4] Panahi, Omid. "Deep learning in diagnostics." *Journal of Medical Discoveries* 2, no. 1 (2025): 1-6.

[5] Yang, Yuntao, Xu Zuo, Avisha Das, Hua Xu, and Wenjin Zheng. "Representation learning of biological concepts: a systematic review." *Current Bioinformatics* 19, no. 1 (2024): 61-72.

[6] Capper, David, David TW Jones, Martin Sill, Volker Hovestadt, Daniel Schrimpf, Dominik Sturm, Christian Koelsche *et al.* "DNA methylation-based classification of central nervous system tumours." *Nature* 555, no. 7697 (2018): 469-474.

[7] Koelsche, Christian, Daniel Schrimpf, Damian Stichel, Martin Sill, Felix Sahm, David E. Reuss, Mirjam Blattner *et al.* "Sarcoma classification by DNA methylation profiling." *Nature communications* 12, no. 1 (2021): 498.

[8] Benfatto, Salvatore, Martin Sill, David TW Jones, Stefan M. Pfister, Felix Sahm, Andreas von Deimling, David Capper, and Volker Hovestadt. "Explainable artificial intelligence of DNA methylation-based brain tumor diagnostics." *Nature Communications* 16, no. 1 (2025): 1787.

[9] Lebrun, Laetitia, Nathalie Gilis, Manon Dausort, Chloé Gillard, Stefan Rusu, Karim Slimani, Olivier De Witte *et al.* "Diagnostic impact of DNA methylation classification in adult and pediatric CNS tumors." *Scientific Reports* 15, no. 1 (2025): 2857.

[10] Gwon K, Yoo J. Out-of-distribution (OOD) detection and generalization improved by augmenting adversarial Mixup samples. Electronics 2023;12:1421. https://doi.org/10.3390/ electronics12061421

[11] Yang, Jingkang, Kaiyang Zhou, Yixuan Li, and Ziwei Liu. "Generalized out-of-distribution detection: A survey." *International Journal of Computer Vision* 132, no. 12 (2024): 5635-5662.

[12] Todeschini, R., Ballabio, D., Consonni, V., Sahigara, F., and Filzmoser, P. "Locally centred Mahalanobis distance: a new distance measure with salient features towards outlier detection." Analytica Chimica Acta 2013.

[13] Hendrycks, D. and Gimpel, K. "A baseline for detecting misclassified and out-of-distribution examples in neural networks." arXiv:1610.02136. ICLR 2017.

[14] Haroush, M., Frostig, T., Heller, R., and Soudry, D. "A Statistical Framework for Efficient Out of Distribution Detection in Deep Neural Networks." ICLR 2022.

[15] Haroush, Matan, Tzviel Frostig, Ruth Heller, and Daniel Soudry. "A Statistical Framework for Efficient Out of Distribution Detection in Deep Neural Networks." In *International Conference on Learning Representations*. 2021.

[16] Sun, Yiyou, Yifei Ming, Xiaojin Zhu, and Yixuan Li. "Out-of-distribution Detection with Deep Nearest Neighbors." *arXiv preprint arXiv:2204.06507* (2022).

[17] Zaeemzadeh, Alireza, Niccolo Bisagno, Zeno Sambugaro, Nicola Conci, Nazanin Rahnavard, and Mubarak Shah. "Out-of-distribution detection using union of 1-dimensional subspaces." In *Proceedings of the IEEE/CVF Conference on Computer Vision and Pattern Recognition*, pp. 9452-9461. 2021.

[18] Liu, W., Wang, X., Owens, J., and Li, Y. "Energy-based out-of-distribution detection." Advances in NeurIPS 33 (2020): 21464–21475.

[19]Mou, Yutao, Keqing He, Pei Wang, Yanan Wu, Jingang Wang, Wei Wu, and Weiran Xu. "Watch the neighbors: A unified k-nearest neighbor contrastive learning framework for OOD intent discovery." In *Proceedings of the 2022 Conference on Empirical Methods in Natural Language Processing*, pp. 1517-1529. 2022.

[20]Hong, Zesheng, Yubiao Yue, Yubin Chen, Lele Cong, Huanjie Lin, Yuanmei Luo, Mini Han Wang *et al.* "Out-of-distribution detection in medical image analysis: A survey." *arXiv preprint arXiv:2404.18279* (2024).

[21]Tran, Quynh T., Alex Breuer, Tong Lin, Ruth Tatevossian, Sariah J. Allen, Michael Clay, Larissa V. Furtado *et al.* "Comparison of DNA methylation based classification models for precision diagnostics of central nervous system tumors." *NPJ Precision Oncology* 8, no. 1 (2024): 218.

[22]Kamoi, Ryo, and Kei Kobayashi. "Why is the mahalanobis distance effective for anomaly detection?." *arXiv preprint arXiv:2003.00402* (2020).

[23]Sehwag, Vikash, Mung Chiang, and Prateek Mittal. "Ssd: A unified framework for self-supervised outlier detection." *arXiv preprint arXiv:2103.12051* (2021).

[24]Pedregosa, Fabian, Gaël Varoquaux, Alexandre Gramfort, Vincent Michel, Bertrand Thirion, Olivier Grisel, Mathieu Blondel *et al.* "Scikit-learn: Machine learning in Python." *the Journal of machine Learning research* 12 (2011): 2825-2830.

[25]Patil, Pratik, Jin-Hong Du, and Ryan J. Tibshirani. "Optimal ridge regularization for out-of-distribution prediction." *arXiv preprint arXiv:2404.01233* (2024).

[26]Virtanen, Pauli, Ralf Gommers, Travis E. Oliphant, Matt Haberland, Tyler Reddy, David Cournapeau, Evgeni Burovski *et al.* "SciPy 1.0: fundamental algorithms for scientific computing in Python." *Nature methods* 17, no. 3 (2020): 261-272.

[27]Tran, Q.T., Breuer, A., Lin, T., Tatevossian, R., Allen, S.J., Clay, M., Furtado, L.V., Chen, M., Hedges, D., Michael, T. and Robinson, G., 2024. Comparison of DNA methylation based classification models for precision diagnostics of central nervous system tumors. *NPJ Precision Oncology*, *8*(1), p.218.

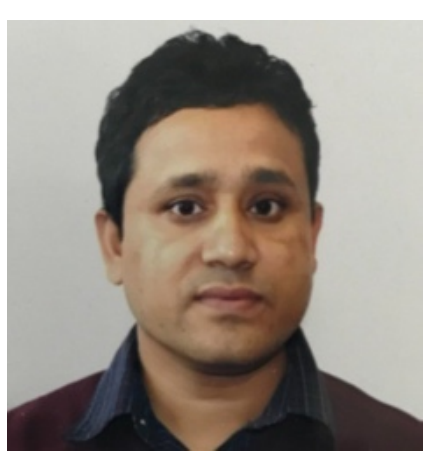

**Dr. Zahangir Alom,** is a senior bioinformatics scientist with extensive experience in computer engineering and biomedical informatics. He received his B.Sc. (Hons.) in Computer Science and Engineering from the University of Rajshahi, Bangladesh, in 2008. He later completed an M.Eng. in Computer Engineering from Chonbuk National University, South Korea, and a master's in clinical Investigation (MSCI) from St. Jude Graduate School of Biomedical Sciences in Memphis, Tennessee, USA. Dr. Alom earned his Ph.D. in Electrical and Computer Engineering from the University of Dayton, Ohio, USA. His research interests include machine learning, deep learning, biomedical informatics, and clinical informatics.

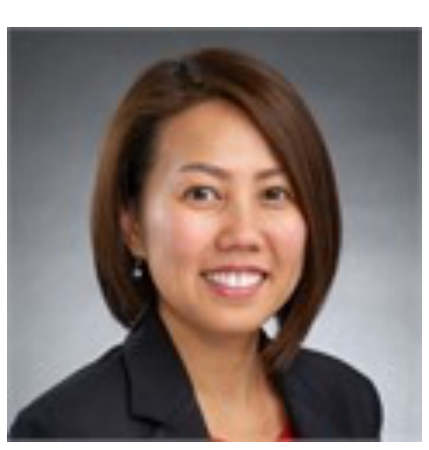

**Dr. Quynh Tran** is a senior bioinformatics scientist with extensive experience in computational biology, systems biology, and translational research. She currently serves as a Senior Bioinformatics Research Scientist at St. Jude Children's Research Hospital, where she has contributed to advancing data-driven approaches in biomedical research since 2018, progressing through roles as Staff Scientist and Bioinformatics Research Scientist. Prior to joining St. Jude, Dr. Tran worked as a Senior Staff Scientist and Systems Biologist at Reynolds American Inc., focusing on integrative biological modeling. She also held an academic position as an Assistant Professor of Preventive Medicine at University of Tennessee Health Science Center, where she contributed to teaching and research in public health and biomedical sciences. Dr. Tran earned both her M.S. and Ph.D. in Statistics, Bioinformatics, and Biology from University of Memphis. She completed her undergraduate studies at Stony Brook University, receiving a B.S. in Computer Science and Applied Mathematics & Statistics. Her work bridges statistical modeling, bioinformatics, and biological systems to support advances in precision medicine and healthcare analytics.

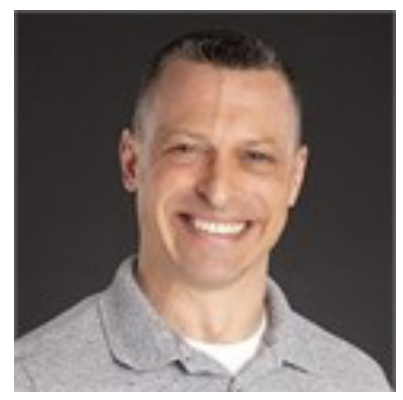

**Dr. Breuer, Alexander** is a Lead Bioinformatics Research Scientist with extensive experience in computational biology and data-driven research. He earned his Ph.D. in Computer Science from Indiana University Bloomington, where he developed a strong foundation in advanced computing and analytical methods. Prior to his doctoral studies, Dr. Breuer completed his bachelor's degree in computer and information science at the University of Delaware. His academic background has supported a career focused on applying computational techniques to complex biological and biomedical problems. In his current role, Dr. Breuer leads bioinformatics research efforts, integrating large-scale data analysis with innovative computational approaches to advance scientific discovery in the life sciences.

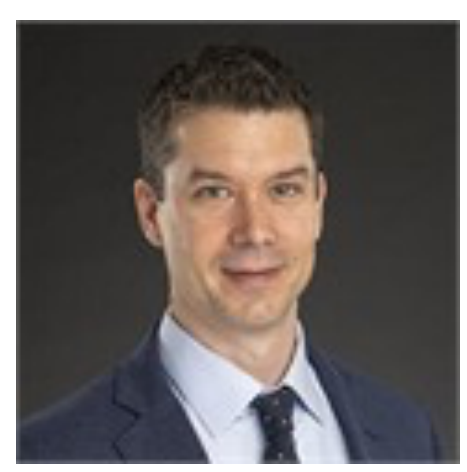

**Brent A. Orr MD, Ph.D.** leads translational pathology research at St. Jude as the Medical Director of Anatomic Pathology and a Member of the St. Jude Faculty. After earning his MD at the University of Illinois College and Medicine and his PhD from the University of Illinois at Urbana-Champaign, Dr. Orr completed his residency in anatomic pathology and a fellowship in neuropathology at The Johns Hopkins Hospital. He uses these clinical and academic experiences as the foundation for his research into molecular tumor subtyping of brain tumors and other solid malignancies**.**